**Framework for engineering of spin defects in hexagonal boron nitride by focused ion beams**

Madeline Hennessey[1,2], Benjamin Whitefield[1,2], Angus Gale[1], Mehran Kianinia[1,2], John A. Scott[1,2,3], Igor Aharonovich[1,2] and Milos Toth[1,2]

[1] School of Mathematical and Physical Sciences, University of Technology Sydney, Ultimo, New South Wales 2007, Australia.
[2] ARC Centre of Excellence for Transformative Meta-Optical Systems, Faculty of Science, University of Technology Sydney, Ultimo, New South Wales 2007, Australia.
[3] The University of Sydney Nano Institute, The University of Sydney, Camperdown, NSW, 2006, Australia

*Hexagonal boron nitride (hBN) is gaining interest as a wide bandgap van der Waals host of optically active spin defects for quantum technologies. Most studies of the spin-photon interface in hBN focus on the negatively charged boron vacancy ($V_B^-$) defect, which is typically fabricated by ion irradiation. However, the applicability and wide deployment of $V_B^-$ defects is limited by $V_B^-$ fabrication methods which lack robustness and reproducibility, particularly when applied to thin flakes (≲10 nm) of hBN. Here we elucidate two key factors that underpin the formation and quenching of $V_B^-$ centers by ion irradiation – density of defects generated in the hBN lattice and recoil-implantation of foreign atoms into hBN. Critically, we show that the latter is extremely efficient at inhibiting the generation of optically-active $V_B^-$ centers. This is significant because foreign atoms such as carbon are commonplace on both the top and bottom surfaces of hBN during ion irradiation, in the form of hydrocarbon contaminants, polymer residues from hBN transfer methods, protective capping layers and substrates. Recoil implantation must be accounted for when selecting ion beam parameters such as ion mass, energy, fluence and incidence angle, which we discuss in the context of a framework for $V_B^-$ generation by high-resolution focused ion beam (FIB) systems.*

**Introduction**

Hexagonal boron nitride (hBN) has emerged as a compelling, wide bandgap host of quantum emitters, and a promising platform for integrated quantum photonics[1–3]. In particular, reports of optically active spin defects have heightened interest in hBN for quantum information processing and quantum sensing applications enabled by a spin-photon interface[4–17]. The most studied spin defect in hBN is the negatively charged boron vacancy ($V_B^-$), shown schematically in Fig. 1(a). The $V_B^-$ center has photoluminescence (PL) and optically detected magnetic resonance (ODMR) spectra characterized by a broad red emission at ~800 nm and a triplet ground state with a zero-field splitting of ~3.5 GHz (Fig. 1(b,c)), respectively.

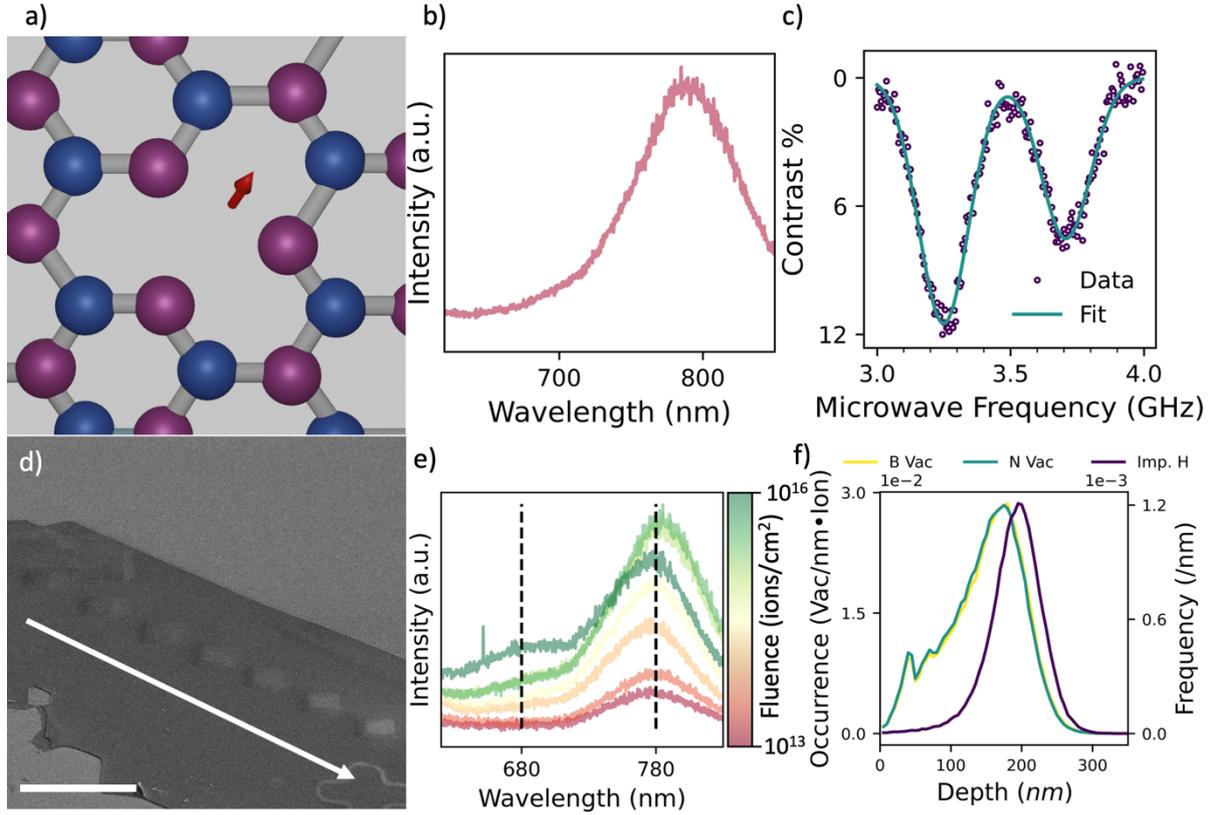

***Figure 1. FIB fabrication of $V_B^-$ centers in hBN.*** *(a) Atomic structure of the $V_B^-$ defect. (b) Typical PL spectrum from an ensemble of $V_B^-$ defects. (c) Typical ODMR spectrum of a $V_B^-$ ensemble acquired under an applied magnetic field of 7 mT. (d) SEM image of a thick (~400 nm) flake of hBN containing 7 regions that had been irradiated by a 15 keV hydrogen ion beam using ion fluences in the range $10^{13}$ - $10^{16}$ $cm^{-2}$. The arrow points from the region of lowest fluence to that of highest fluence. Scale bar = 50 μm. (e) PL spectra versus ion fluence acquired from the 7 regions shown in d. Each spectrum contains the $V_B^-$ emission at ~780 nm and a background emission at ~680 nm. (f) Simulated depth distributions of the generation rates of boron and nitrogen monovacancies (left axis) and hydrogen ions implanted into bulk hBN (right axis). The simulations were performed using 15 keV $H^+$ ions.*

Interest in spin defects hosted by hBN stems primarily from the van der Waals nature of hBN, which can be exfoliated into layers that are stable and easy to integrate in devices and photonic nanostructures[18–22]. However, widescale deployment of applications based on spin defects in hBN requires a robust $V_B^-$ fabrication technique – ideally, one that is site-specific and scalable. $V_B^-$ defects can be fabricated by various methods, including irradiation by femtosecond laser pulses[23,24], MeV electrons[25–27], keV ions[28,29] and neutrons[8,30–32]. Of these, keV ions are appealing because the approach is scalable, and high-resolution focused ion beam (FIB) tools are both broadly available and offer fine control over lateral and depth distributions of $V_B^-$ defects in hBN. However, an outstanding challenge is a lack of reproducibility and robustness. Here, we address this issue and elucidate processes that underpin the generation and quenching of optically active $V_B^-$ defects during ion irradiation – namely, the density of defects generated in the hBN lattice, and implantation of foreign atoms into hBN. Critically, the latter include unintended recoil implantation of impurities into hBN. Specifically, we show that recoil implantation of carbon present on the surface of hBN causes quenching of $V_B^-$ fluorescence even when very light (hydrogen) ions are employed. We use this insight to discuss the roles ion beam parameters in the generation of $V_B^-$ centers, and develop a framework for FIB fabrication of $V_B^-$ centers in hBN by ion irradiation.

**Results and discussion**

Defect generation in hBN by ion impact has been explored previously using modelling studies focused on atomic structures of the defects and defect formation kinetics[33,34]. These studies are, however, insufficient to explain the roles of ion beam parameters on $V_B^-$ fluorescence. We therefore start with two ion irradiation experiments designed to elucidate the primary $V_B^-$ fluorescence generation and quenching mechanisms.

Fig. 1(d) is a scanning electron microscope (SEM) image of a thick (~400 nm) flake of hBN with 7 rectangular regions that had been irradiated by a 15 keV hydrogen ion beam as a function of ion fluence. PL spectra collected from the 7 regions are shown in Fig. 1(e). Each spectrum consists of two peaks indicated on the figure – a characteristic $V_B^-$ emission at approximately 800 nm, and an overlapping background emission from other (unidentified) defects generated by ion impacts. The intensity of the background PL emission increases with ion fluence. It is undesirable as it overlaps with the $V_B^-$ peak, and does not allow spin manipulation or readout. Our objective is to maximize the $V_B^-$ emission and minimize the background emission (noting that annealing cannot be used after ion irradiation as it causes quenching of the $V_B^-$ emission[35]). The intensity of the $V_B^-$ emission has a maximum at an ion fluence of ~5 x $10^{15}$ cm$^{-2}$. The initial increase in $V_B^-$ PL intensity with fluence corresponds to an increase in the concentration of $V_B^-$ centres. We attribute the subsequent decrease to two processes. The first is formation of defects other than $V_B^-$ by ion impact, which introduce nonradiative relaxation pathways or act as charge traps that donate/accept electrons to/from $V_B^-$ centres. The second is restructuring of $V_B^-$ centres by ion impact (e.g., conversion of boron monovacancies to divacancies and other defect clusters). These processes are negligible at low fluence, but become significant and eventually dominate as defect densities increase with ion fluence. Hence, there exists an optimum fluence at which the intensity of the $V_B^-$ PL emission is maximized, as is seen in Fig. 1(e).

To assist with interpretation of ion irradiation experiments, we performed simulations of ion-solid interactions using the software package Stopping and Range of Ions in Matter (SRIM)[36]. Fig. 1(f) shows simulated depth distributions of the generation rates of boron monovacancies in hBN. It also shows the distributions of nitrogen monovacancies and implanted hydrogen atoms, which are representative of native defects and impurities that can potentially inhibit $V_B^-$ fluorescence. The simulations were performed using 15 keV $H^+$ ions and a semi-infinite slab of bulk hBN (see Methods for details). The depth distributions of the B and N monovacancies are approximately the same and overlap partially with that of the implanted H atoms. The ion range scales with energy (not shown) and the simulations can be used to select the ion beam energy needed to confine/transmit ions to/through a flake of hBN.

The behaviour seen in Fig. 1(e) should, in principle, be robust and reproducible. That is, there should be an optimum ion fluence for the generation of $V_B^-$ centers, determined by ion beam parameters such as ion mass, energy and fluence which affect the concentrations and spatial distributions of defects and implanted atoms in hBN. However, in practice, this is not the case. The curve of $V_B^-$ emission intensity versus ion fluence often varies substantially from flake to flake, particularly when the hBN flakes are thin. The problem persists even when the flakes are exfoliated from a single crystal of hBN and both the flake thickness and ion irradiation conditions are fixed. We propose that it is caused by recoil implantation of foreign atoms into hBN, which gives rise to quenching of the $V_B^-$ emission. Foreign atoms are, typically, present on both the top and bottom surfaces of hBN flakes in the form of hydrocarbon contaminants and, in some cases, polymer residues from hBN exfoliation and transfer methods, or capping layers such as graphene electrodes and/or substrates. The quantities of residues and contaminants often vary between hBN flakes, and can, in principle, cause the poor reproducibility of $V_B^-$ generation recipes. To demonstrate that recoil implantation does indeed quench $V_B^-$ centers, we performed the experiment detailed in Fig. 2.

A 38 nm thick flake of hBN was partially capped with 7 nm multilayer graphene (MLG) to produce "capped" and "uncapped" areas, both of which were then irradiated with hydrogen ions (10 keV, $5\times10^{14}$ ions/cm$^2$) to generate $V_B^-$ centers. The uncapped region was then partially covered with a second (22 nm) flake of MLG to produce a "control" region for PL measurements, as is shown in Fig. 2(a). PL spectra were then collected from all three regions of hBN. The uncapped region of hBN exhibits a relatively intense $V_B^-$ emission, and a weak background emission (black spectrum labelled "Uncapped" in Fig 2(b)). The "capped" region that was irradiated through a 7 nm MLG layer exhibits a relatively weak $V_B^-$ emission and a relatively intense background emission (green spectrum labelled "Capped" in Fig 2(b)). A PL spectrum from the control region (Fig. 2b, blue) shows that the bulk of the observed reduction in $V_B^-$ emission in the region capped by 7 nm MLG is not caused by absorption of PL by the MLG (the weak sharp peaks below 700 nm are Raman lines from the 22 nm MLG layer). To quantify the quenching, we fit each spectrum using two Gaussians, one each for the $V_B^-$ and the background emission (the fitting was done in the energy domain). The peaks used to fit the $V_B^-$ emission are shown in Fig 2(b). In the control region, the $V_B^-$ intensity is reduced by ~17% due to PL absorption in the 22 nm MLG layer. In the capped region, the $V_B^-$ intensity is reduced by ~50%. We attribute the bulk of this reduction to quenching of $V_B^-$ centers by recoil implantation of C into hBN by the hydrogen ions.

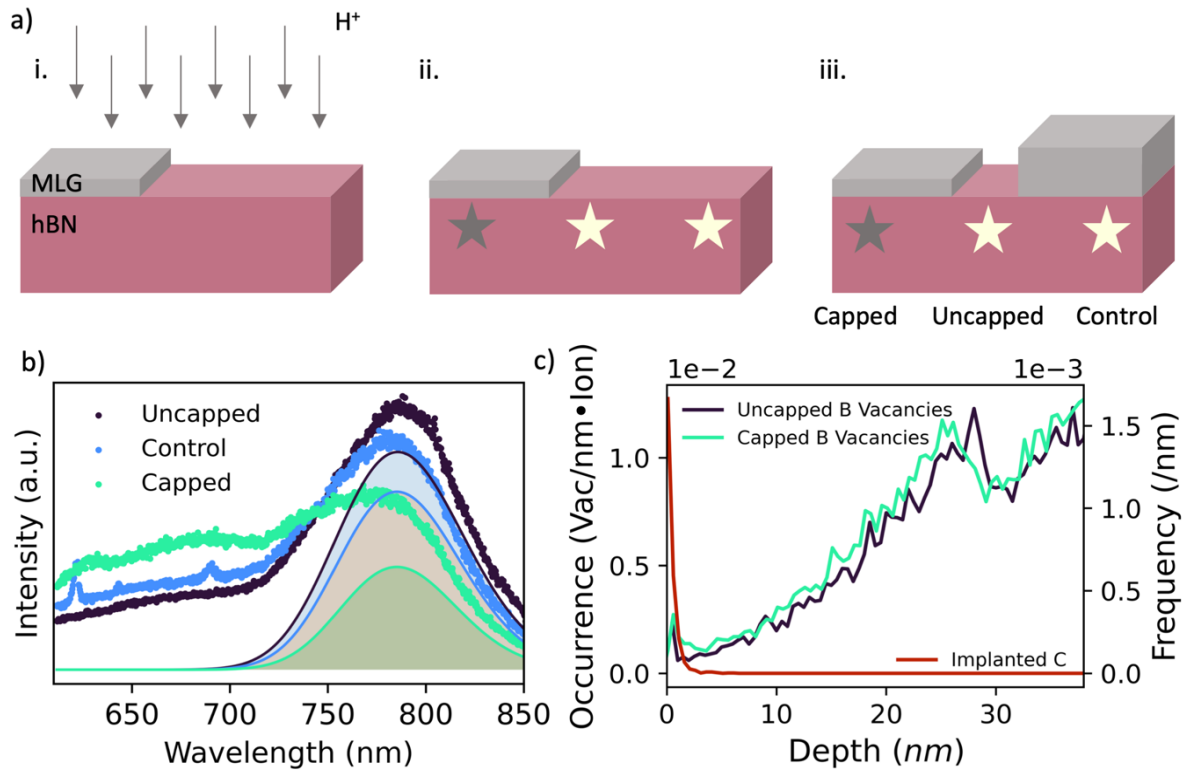

*Figure 2. Quenching of $V_B^-$ centers by recoil implantation of carbon into hBN.* (a) Schematic illustration of the sample processing sequence: [i] a 38 nm thick flake of hBN was partially capped with 7 nm MLG, and subsequently irradiated with hydrogen ions (10 keV, $5\times10^{14}$ cm$^{-2}$); [ii] $V_B^-$ centers were generated by the ions in both the capped and uncapped regions of hBN (grey and yellow stars, respectively); [iii] 22 nm MLG was transferred onto the hBN flake to create a control region for PL measurements. (b) PL spectra collected from the three regions of hBN: uncapped hBN (black points), hBN that was capped with 7 nm MLG before ion irradiation (green), and a control region that was covered with 22 nm MLG after ion irradiation (blue). Also shown are corresponding fits of the $V_B^-$ emission (black, green and blue lines, respectively). (c) Simulated depth distributions of boron monovacancies generated in the capped hBN (green curve, left axis) and carbon atoms recoil-

implanted from MLG into hBN (red curve, right axis). The simulations were performed using 38 nm hBN capped with 7 nm MLG, and 10 keV $H^+$ ions. The bottom axis is the depth below the hBN surface (the MLG region spans -7 to 0 nm, and is omitted from the plot). Also shown is the distribution of boron monovacancies simulated using the same ion beam conditions and uncapped 38 nm hBN (black curve, left axis).

To substantiate the claim of recoil implantation, we performed the SRIM simulations shown in Fig. 2(c). The simulated generation rates and depth distributions of boron monovacancies by 10 keV $H^+$ ions in 38 nm hBN are almost identical in uncapped hBN and in hBN capped with 7 nm MLG (black and green curves in Fig. 2(c), respectively). That is, the ~50% reduction in $V_B^-$ emission observed in Fig. 2(b) cannot be explained by a change in the number of vacancies generated in hBN by the ion beam. The simulations do, however, show that C is indeed recoil-implanted into hBN (red curve in Fig. 2(c)). The simulated carbon implantation is caused by energy and momentum transfer from 10 keV $H^+$ ions to C atoms, and it occurs despite the very low mass of $H^+$.

We note that, in Fig. 2(c), the implanted C atoms are confined to the top few nanometers of the 38 nm hBN layer. This simulated depth profile, however, does not account for channelling effects[36]. Channelling causes a small fraction of impurities to be incorporated at depths much greater than predicted by SRIM simulations. Such impurities can cause a dramatic reduction in PL intensity due to the generation of efficient nonradiative recombination pathways. For example, irradiation of diamond by (2 to 30 keV) oxygen ions has been shown previously to quench the nitrogen-vacancy emission at depths that are much greater (i.e., hundreds of nanometers) than oxygen implantation depths predicted by SRIM simulations[37,38]. This is also expected in the present case of recoil implantation of C into hBN. Channelling likely caused C incorporation throughout the 38 nm hBN layer, and thus caused the significant (~50%) quenching of the $V_B^-$ emission observed in Fig. 2(b).

The most pertinent conclusion from the results in Fig. 2 is that the presence of foreign atoms on the surface of hBN during ion irradiation can cause quenching of $V_B^-$ fluorescence. Such impurities can originate in capping layers, as well as contaminants/residues which must be minimised prior to ion irradiation. In addition, if the ion range is greater than the hBN thickness, recoil implantation of atoms located at the bottom hBN surface can occur. As is discussed in detail below, the consequences of recoil implantation are most significant if heavy ions (such as $Ar^+$, $Xe^+$ or $Ga^+$) are employed.

We note that recoil implantation of impurities using commercial FIB instruments has been demonstrated previously. In these studies, it was exploited for optical doping of diamond by recoil implantation of atoms from solid thin films[39,40], as well as surface-adsorbed $N_2$, $NH_3$, and $NF_3$ molecules which were delivered in small quantities (at room temperature) using a gas injector[41].

Having demonstrated experimentally the two key factors that influence the $V_B^-$ emission – namely, the defect concentration (Fig. 1(e)) and knock-on implantation of impurities (Fig. 2(b)) – we will now use SRIM simulations to discuss in detail the role of ion beam parameters in $V_B^-$ generation. We focus on hBN of varying thickness, with an emphasis on flakes thinner than ~10 nm due to their appeal for quantum sensing applications.

First, we consider the generation of boron monovacancies versus ion beam energy and ion beam incidence angle θ. Fig. 3(a) shows a plot of the cumulative boron vacancy generation rate, G (vacancies per incident ion), as a function of depth. The simulations were performed using 10 keV $H^+$ ions, 10 incidence angles between 0 and 89°, and bulk hBN. At each depth, d, the value of G(d) is the mean number of vacancies per ion generated between the hBN surface and d. G(d) therefore corresponds to the net vacancy generation rate in a slab of hBN enclosed by the surface (where d = 0) and d.

Moreover, G(d) also approximates the vacancy generation rate in a suspended hBN flake of thickness d, and trends in dependencies of G(d) on ion parameters such as energy (E) and θ also apply to suspended hBN flakes of varying thickness d. The simulations therefore provide an efficient means to indicate how ion beam parameters affect $V_B$ generation in both the near-surface region of bulk hBN and suspended flakes of hBN.

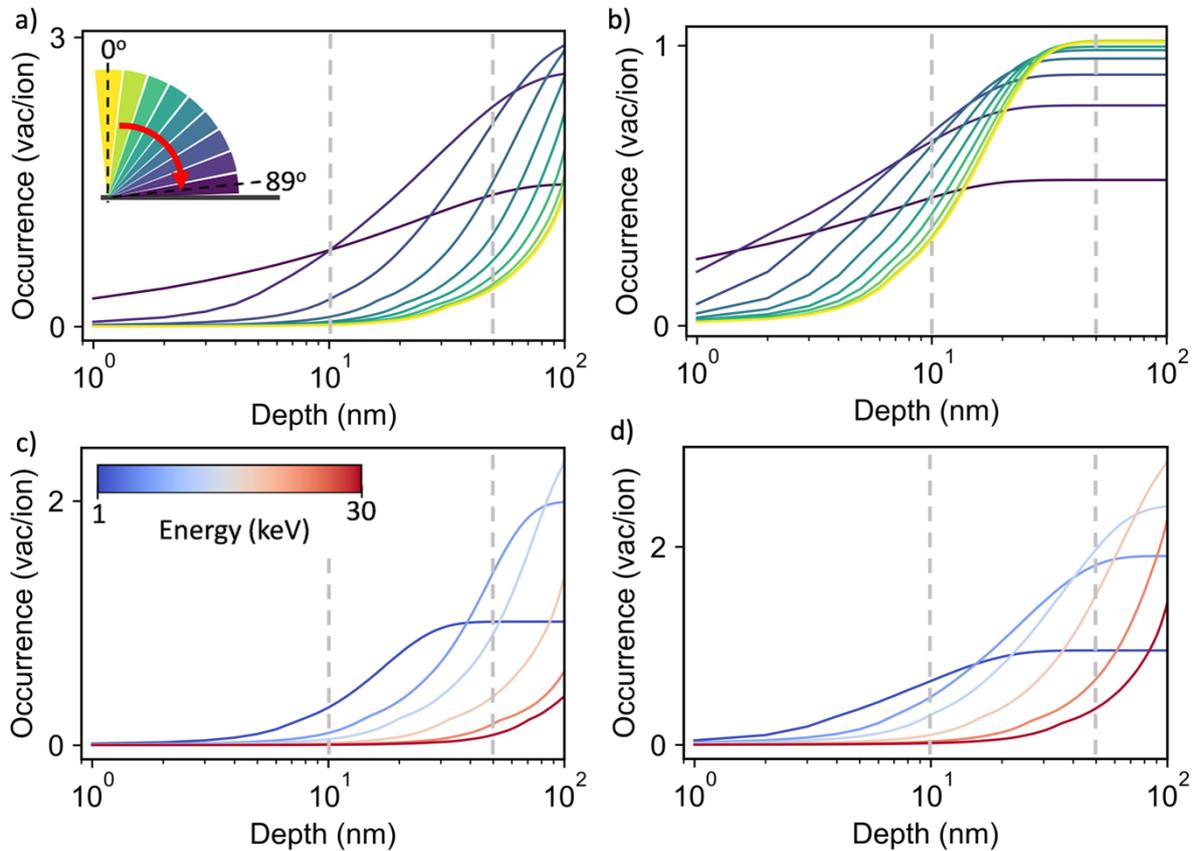

*Figure 3. Boron vacancy generation in boron nitride simulated versus depth. (a) Boron vacancy generation versus incidence angle by 10 keV $H^+$ ions. (b) Boron vacancy generation versus incidence angle by 1 keV $H^+$ ions. (c) Boron vacancy generation versus energy for normal incidence $H^+$ ions. (d) Boron vacancy generation versus energy for $H^+$ ions incident at 60˚.*

Fig. 3(a) shows that, for thin flakes (defined here as < 10 nm) bombarded with 10 keV light ions ($H^+$), G(d) increases with θ for all angles up to ~89°. This trend also holds at higher ion energies (not shown), but changes at low energies. An example is shown in Fig. 3(b) where, at 1 keV, G(d) has a maximum at the "optimal" angle $θ_m$ of ~70°. For flakes much thicker than 10 nm, $θ_m$ varies with energy (e.g., at 50 nm, $θ_m$ ~80°, as is seen in Fig. 3(a)). The general trend is that $θ_m$ decreases with increasing flake thickness and with decreasing ion beam energy. This is illustrated in Fig. 3(c) and (d) by plots of G(d) at six energies between 1 and 30 keV, and ion incidence angles of 0 and 60°, respectively. The data can be summarized succinctly by constraining the energy and incidence angle to the ranges of 1 - 30 keV and 0 - 60°, which are readily accessible by commercial FIB instruments. Within these constraints, the $V_B$ generation rate per light ion incident onto a thin suspended flake of hBN is maximized by minimizing the ion energy and maximizing the incidence angle. The reciprocal scaling with energy is a consequence of the fact that the ion range increases with energy, which we discuss further below, in the context of sputtering and recoil implantation.

We note, that the trends seen in Fig. 3 correlate with the frequency of ion-atom collisions in boron nitride. They are therefore also applicable to the generation of (undesirable) nitrogen vacancies, interstitials and larger defect clusters by the ions. The concentration of such defects increases with fluence and they can, in principle, interact with and quench the fluorescence of $V_B^-$ defects. Hence, at any given ion energy and incidence angle, there exists an ion fluence that is optimal for the generation of optically active $V_B^-$ defects (as we have shown experimentally in Fig. 1(e)).

Next, we turn to ion beam sputtering which causes thinning of hBN flakes. This is undesirable and must be minimized, particularly when thin flakes of hBN are employed. Moreover, simulations of sputtering by keV ions are insightful because both sputtering and recoil implantation are influenced by the amount of energy transferred from ions to surface atoms – that is, simulations of sputtering provide useful insights into recoil implantation.

Fig. 4(a) and (b) show plots of the sputter yield (Y) versus E (1 to 30 keV) and θ (0 to 89°), respectively, for $H^+$ ions. In general, Y decreases with increasing E (except at very high angles approaching ~89°, where the ion beam is almost parallel to the hBN) and increases with θ, mainly because the ion range increases with energy and decreases with θ. That is, the greater the range, the smaller the amount of energy that is transferred to surface atoms per ion and the lower the sputter yield. Hence, to minimise sputtering by light keV ions, the energy should be maximized and the incidence angle should be minimized. These trends are also observed with other light keV ions such as $H_3^+$ and $He^+$ (Fig. 5(a,b)). However, the energy dependence of Y is inverted for heavy ions. More specifically, Y increases with E for heavy keV ions such as $Ar^+$ and $Xe^+$, as is shown in Fig. 5(a), because the ion range scales inversely with ion mass, and an increase in the energy of keV heavy ions results in an increase in the amount of energy deposited (per ion) to surface atoms.

Critically, the above argument also applies to recoil implantation – that is, to minimize recoil implantation by light keV ions, the ion energy should be maximised, but for heavy keV ions the energy should be minimized.

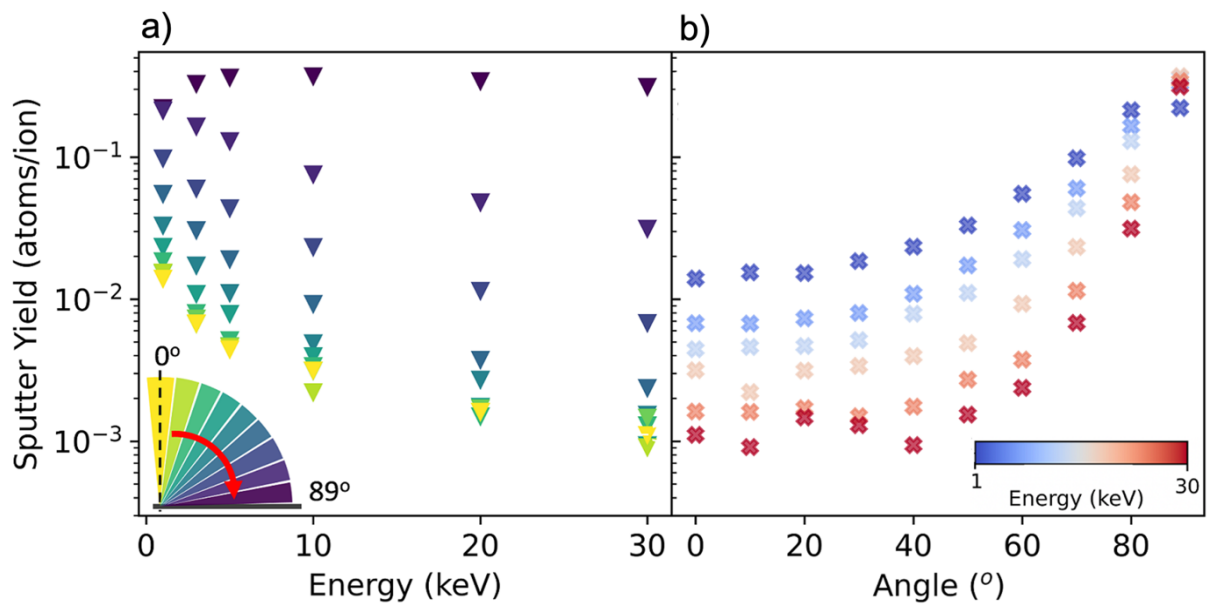

*Figure 4. Simulated sputter yield of boron nitride irradiated by $H^+$ ions. (a) Sputter yield versus energy at 10 incidence angles between 0° and 89°. (b) Sputter yield versus incidence angle at 6 ion energies between 1 and 30 keV.*

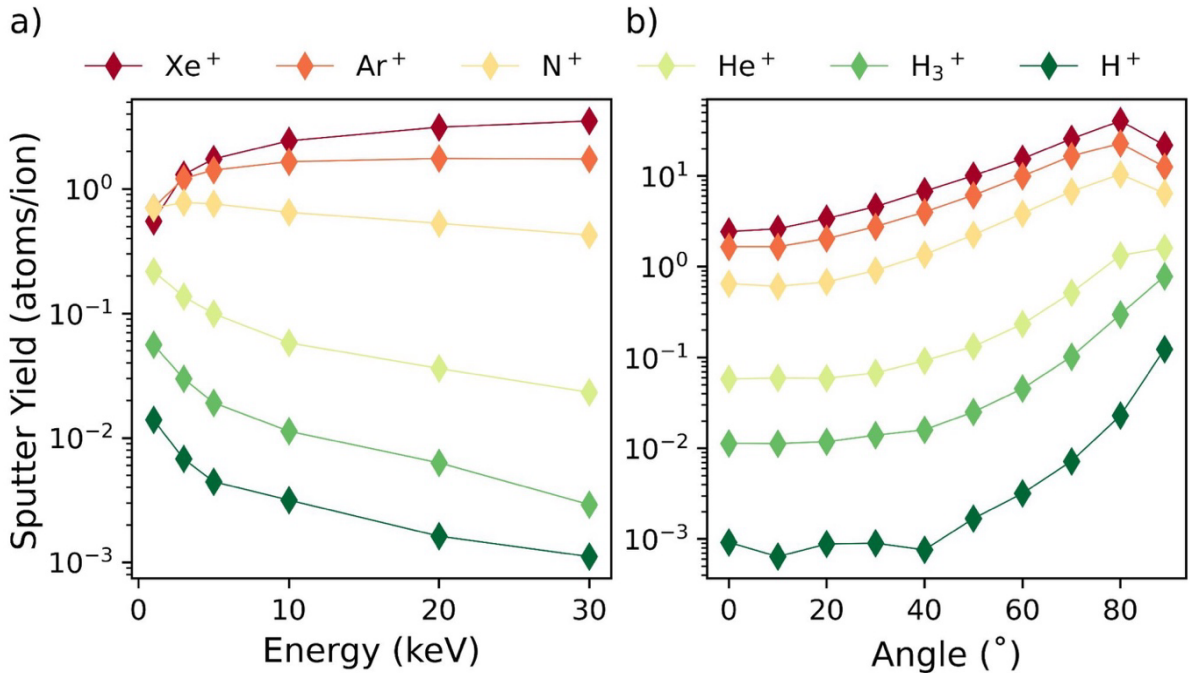

*Figure 5: Simulated sputter yield of boron nitride irradiated by light and heavy ions.* (a) Sputter yield versus energy for normal incidence $Xe^+$, $Ar^+$, $N^+$, $He^+$, $H_3^+$ and $H^+$ ions. (b) Sputter yield versus incidence angle for 10 keV $Xe^+$, $Ar^+$, $N^+$, $He^+$, $H_3^+$ and $H^+$ ions.

The data in Fig. 5(a) illustrate that, quantitatively, the absolute sputter yield for heavy ions such as $Xe^+$ is much greater than for light ions. That is, the mass of keV ions is more important than the ion energy, and the mass should be minimized to minimize both sputtering and recoil implantation.

For completeness, the angular dependence of the sputter yield, $Y(\theta)$, by heavy ions is shown in Fig. 5(b). It is qualitatively similar to light ions – Y increases with θ, except at angles greater than ~80˚. Such large incidence angles are, however, impractical as they correspond to the situation where the ion beam is almost parallel to the hBN surface, and the beam placement accuracy is therefore compromised. Hence, the general conclusion is the same as for light ions – to minimize sputtering, the ion incidence angle should be minimized. Recoil implantation is, in this instance, more nuanced and case-specific because the incidence angle also affects channeling (which is not accounted for by SRIM).

Using the above results, we can now construct a general framework for $V_B^-$ fabrication by focused ion beams. We start by noting that a universal recipe that is optimal in all circumstances does not exist due to application-specific variables such as hBN thickness and the presence of foreign atoms above/below the hBN. We therefore provide general guidelines that can be adapted for specific use-cases.

**Recoil implantation (intermixing)** introduces impurities (i.e., extrinsic defects) into hBN and must be minimized. These impurities can originate at surface contaminants, residues, substrates or device components such as electrodes at the top and bottom surfaces of hBN. Ideally, all impurities should be eliminated by using clean, suspended flakes of hBN. If impurities are unavoidable, then the ion beam parameters should be selected to minimize the amount of energy deposited (per ion) near hBN interfaces (i.e., the top and bottom surfaces of a flake of hBN). Based on our observations (Fig. 2), intermixing is very effective at suppressing $V_B^-$ fluorescence even when very light (hydrogen) ions are employed and it must be minimized as a matter of priority in any recipe designed to optimize the fabrication of $V_B^-$ centers.

**Ion implantation** should, likewise, be minimized. This can be achieved by transmitting the beam through hBN by minimizing the ion mass, maximizing the ion energy, and by employing suspended, thin flakes of hBN.

**The sputter yield** should be minimized. The mass of keV ions is the key parameter for sputtering, and it should be minimized.

**The ion FIB spatial resolution** (i.e., beam diameter, beam tails and beam placement accuracy) must be optimized for some applications of $V_B^-$ centers. Hence, low ion beam energies (below ~5 keV) and very high incidence angles must be avoided, as they exacerbate ion optical aberrations[42], and beam placement accuracy, respectively.

The above can be distilled into the following simple framework for the generation of $V_B^-$ centers by keV focused ion beams. Recoil implantation of impurities should be avoided by employing bare, clean flakes of hBN, which should ideally be suspended to avoid ion beam intermixing of hBN and a substrate. The ion mass should be minimized and the optimal ion fluence must be determined experimentally using a calibration curve of $V_B^-$ PL intensity versus fluence (i.e., ion dose per unit area).

## Conclusion

We have developed a practical framework for the fabrication of $V_B^-$ defects in hBN using focused ion beams. We identified processes that lead to the generation and quenching of $V_B^-$ centers, and presented guidelines for the design of robust $V_B^-$ engineering protocols. Our findings will accelerate applications based on quantum sensing using $V_B^-$ centres in hBN.

## Acknowledgements

The authors acknowledge financial support from the Australian Research Council (CE200100010, FT220100053) and the Office of Naval Research Global (N62909-22-1-2028) for financial support. The authors also thank the UTS ANFF Optofab for access to facilities.

## Methods

Flakes of hBN were mechanically exfoliated onto a 285 nm silicon oxide substrate using scotch tape. The substrate was then annealed at 300°C for 2 hours and atomic force microscopy (AFM) was used to measure flake thickness.

Multilayer graphene (MLG) (thickness ~7 nm, pre-characterized by AFM) was transferred onto a 38 nm hBN flake using a PDMS stamp under an optical microscope. Ion beam processing was performed using a Thermo Fisher plasma-source FIB/SEM dual-beam. The sample was irradiated with hydrogen (10 keV, $5 \times 10^{14}$ cm$^{-2}$) at normal incidence. A second 22 nm MLG flake was transferred via the same technique.

PL spectroscopy was performed using a home-made confocal PL setup described elsewhere[43], using a 532 nm laser, an excitation power of 5 mW, and an integration time of 120 seconds and averaged over 3 scans. ODMR was performed using a second confocal setup[8] using a 3.3 mW, 532 nm laser.

Ion-solid interactions were simulated using the SRIM package[36]. Each simulation was performed using 99,999 incident ions using the ion species, energies and incidence angles specified in the figures.